\documentclass[prb,notitlepage,twocolumn]{revtex4}
\usepackage{amsmath}
\usepackage{amssymb}
\usepackage{bm}
\usepackage{graphicx}
\usepackage{times}

\begin{document}

\title{Inverse dispersion method for calculation of complex photonic band diagram and $\cal{PT}$-symmetry.}

\author{Mikhail~V.~Rybin${}^{1,2}$}
\email{m.rybin@mail.ioffe.ru}
\author{Mikhail~F.~Limonov${}^{1,2}$}

\affiliation{$^1$ Ioffe Physical-Technical Institute, St.~Petersburg 194021, Russia\\
$^2$Department 
of Nanophotonics and Metamaterials, ITMO University, St.~Petersburg 197101, Russia
}

\begin{abstract}
We suggest an inverse dispersion method for calculating photonic band diagram 
for materials with arbitrary frequency-dependent dielectric functions. The 
method is able to calculate the complex wave vector for a given frequency by 
solving the eigenvalue problem with a non-Hermitian operator. The analogy with 
$\cal{PT}$-symmetric Hamiltonians reveals that the operator corresponds to 
the momentum as a physical quantity and the singularities at the band edges 
are related to the branch points and responses for the features on the band 
edges. The method is realized using plane wave expansion technique for 
two-dimensional periodical structure in the case of TE- and TM-polarization. 
We illustrate the applicability of the method by calculation of the photonic band diagrams of 
an infinite two-dimension square lattice composed of dielectric cylinders using the 
measured frequency dependent dielectric functions of different materials 
(amorphous hydrogenated carbon, silicon, and chalcogenide glass). We show 
that the method allows to distinguish unambiguously between Bragg and Mie 
gaps in the spectra.
\end{abstract}

\date{\today}


\maketitle

\section{Introduction}

In the past decades, a significant progress has been achieved in the fabrication of light-controlling nanostructures including metamaterials \cite
{enghata2006electromagnetic, smith2010metamaterials} and photonic crystals 
\cite{g101,g142}. Hence the ability to accurately calculate photonic 
properties is required for designing novel devices, while the photonic band 
diagram, i.e., frequency $\omega$ vs. wave vector $\mathbf{k}$, being the most 
important inherent property of periodic media \cite{g101}. The 
common approach to the problem consists in reducing Maxwell's equations to the 
standard eigenproblem for the frequency as an eigenvalue \cite 
{g401,wang1993multiple,g404,g405,moroz2011multiple}. The first successful 
approach \cite{g401} assumed that the dielectric function does not vary with 
frequency. The authors used the plane wave expansion technique to solve the 
wave equation with a Hermitian operator for the magnetic field. The subsequent papers 
suggested numerical improvements such as application of the fast Fourier 
transform method, simultaneous calculations of a block of eigenvectors, etc
\cite{meade1993accurate,g405}. In the case of a cylindrical or spherical symmetry of 
the scatterer the Green's function method was proposed 
\cite{leung1993multiple,wang1993multiple,moroz1995density}. With these methods the photonic band diagram of the 
most important structures as well as the conditions for the appearance of a complete photonic 
band gap were determined \cite{g404,moroz1999three,rybin2015band}.

At the same time, most materials strongly interacting with light have a
frequency dependence of dielectric function $\varepsilon(\omega)$ and 
accurate methods should take this circumstance into account. A number of approaches suggesting 
that $\varepsilon(\omega)$ is a Lorentz-like single pole function have been 
proposed \cite 
{pendry1996calculating,kuzmiak1997photonic,kuzmiak1998distribution, 
halevi2000tunable,sakoda2001photonic, 
moreno2002band,huang2003field,toader2004photonic,raman2010photonic}. 
The following techniques have been used to calculate the photonic band 
diagram: excluding the frequency from the operator to yield a generalized 
eigenproblem \cite{kuzmiak1997photonic}; 
considering the polarization field and solving it in the time domain \cite
{sakoda2001photonic}; including the mechanical problem for charged 
particles along with the electric and magnetic fields into the operator \cite
{raman2010photonic}.
However, the optical range of electronic interband transition is not 
described correctly by the Lorentz-type functions and other advanced methods should be 
employed. In addition, the traditional methods calculate only propagating modes with 
a real wave vector. They cannot provide any information about the light propagation
within the bandgap frequencies, and about the rate at which the evanescent mode disappears.
Also a study of modes within the forbidden band is important for understanding the transition between two regimes of light propagation in a dielectric periodical structure, i.e. the regime where the propagation of light is mainly influenced by the interference between multiple scattered Bragg waves (photonic crystals), and the regime where the propagation of light is mainly determined by the properties of each single element of the materials (Mie modes of metamaterials). The transition is accompanied by complicated interactions of multiple bands and the correct interpretation is a key to creating a phase diagram photonic crystal/all-dielectric metamaterial \cite{rybin2015phase}. Therefore we need a tool to explore evanescent modes within a bandgap and distinguish whether it is the Bragg gap related to the non-localized Bragg resonance or the Mie gap originated from a localized Mie-like resonance.

An alternative to the conventional techniques is to consider methods for calculating complex wave vector while the frequency is treated as parameter. We can find it from transcendent KKR equation (Green's function method) \cite{leung1993multiple,wang1993multiple,moroz1995density} using so called on-shell approach (the term is borrowed from scattering theory that means solutions on the energy shell \cite{g148}). Another method is based on Wannier functions \cite{hermann2008photonic}.
Using these methods we eliminate two limitation of the direct $\omega(k)$ approaches. The first one is related to the strong frequency dependence of $\varepsilon(\omega)$. For traditional numerical techniques to be applicable, $\omega$ should be excluded from the operator but this can only be done for restricted classes of $\varepsilon(\omega)$ function. The second limitation is that the direct methods are intended for calculating the propagating modes. By contrast, in implementation of the inverse $k(\omega)$ method the wave vector $k$ should be excluded from the operator. Fortunately, in most cases the spatial dispersion $\varepsilon(k)$ is negligible and the dielectric function is independent of the wave vector considered to be the eigenvalue. Also we can study prohibited evanescent modes by setting $\omega$ within the bandgap and finding whether or not the mode originates from the localized resonance and the structure can be 
homogenized \cite{datta1993effective,kirchner1998transport,Tartar2009homogenization}.
However both Green's function method\cite{leung1993multiple,wang1993multiple,moroz1995density} and the Wannier method \cite{hermann2008photonic} are quite complicated techniques relative to straightforward plane wave expansion techniques. On the other hand plane wave based methods (so-called rigorous coupled wave analysis) are widely used for calculating transport properties of periodic slab by fixing tangential component of the wave vector and the frequency \cite{moharam1981rigorous,li1993convergence,lalanne1996highly,shi2005revised,shishkin2014multiple}.

In this paper we propose an alternative to the conventional methods by introducing an inverse dispersion method for calculating the photonic band diagram. We reduce Maxwell's equations to a problem with the eigenvalue $k$, while $\omega$ is considered to be a real parameter. We 
name this method \emph{the inverse $k(\omega)$ problem} instead of \emph{the direct $\omega(k)$ problem}. These method has straightforward implementations in plane wave basis as well as in real space. 
By way of illustration, we 
consider three periodic structures with the same geometry but different materials, namely
amorphous hydrogenated carbon a-C:H (low dielectric contrast with air $\varepsilon\approx 4$ in VIS), silicon ($\varepsilon\approx 16$ in VIS), and chalcogenide glass Ge$_2$Sb$_2$Te$_5$ (high dielectric contrast $\varepsilon\approx 30$ in IR). We note that the 2D square lattice of dielectric cylinders acts as a photonic crystal at a low dielectric contrast \cite{quinonez2006band}, whereas in a special structural condition the same structure with a high enough contrast behaves as a metamaterial with $\mu<0$ \cite{o2002photonic}. In the present study we use dielectric function $\varepsilon(\omega)$ measured  in Refs. \onlinecite
{li1980refractive,aspnes1983dielectric,palik1998handbook,nvemec2009ge} to calculate the photonic band diagram.


\section{Inverse problem}

In this section we formulate the electrodynamic equations in the form suitable for the inverse dispersion approach. In general, the electromagnetic field is described by the Maxwell's equations which for the frequency harmonics $\exp(-i\omega{}t)$ read
\begin{subequations}
\begin{eqnarray}
\nabla\times\mathbf{E}=i\omega\mu_{0}\mu(\mathbf{r})\mathbf{H} ,
\\
\nabla\times\mathbf{H}=-i\omega\varepsilon_{0}\varepsilon(\mathbf{r})\mathbf{E}.
\label{eq:LorPlusBg}
\end{eqnarray}
\label{eq:Maxwell}
\end{subequations}
Here $\mathbf{E}$ and $\mathbf{H}$ are respectively the electric and magnetic vectors, 
$\varepsilon(\mathbf{r})$ and $\mu(\mathbf{r})$ are the dielectric permittivity and the magnetic permeability. In periodic structures the solutions can be written as the Bloch waves $\mathbf{E}(\mathbf{r})={\cal E}_{\mathbf{k}}(\mathbf{r})e{}^{i\mathbf{k}\mathbf{r}}$
and $\mathbf{H}(\mathbf{r})={\cal H}_{\mathbf{k}}(\mathbf{r})e{}^{i\mathbf{k}\mathbf{r}}$.
By substituting these into Eqs.~(\ref{eq:Maxwell}) we obtain the generalized
eigenproblem for the frequency 
\begin{equation}
\left(\nabla+i\mathbf{k}\right)\times
\left(\begin{array}{c}
{\cal E}\\
{\cal H}
\end{array}\right)=i\omega\left(\begin{array}{cc}
0 & \mu_{0}\mu\\
-\varepsilon_{0}\varepsilon & 0
\end{array}\right)\left(\begin{array}{c}
{\cal E}\\
{\cal H}
\end{array}\right).
\end{equation}
Instead of applying the operator at r.h.s. once again to get the direct problem 
we extract the wave vector from the operator at l.h.s. and move
the matrix with the frequency to l.h.s. Now, we obtained generalized eigenproblem
for the wave vector
\begin{equation}
\left(\begin{array}{cc}
i\nabla\times & \omega\mu_{0}\mu\\
-\omega\varepsilon_{0}\varepsilon & i\nabla\times
\end{array}\right)\left(\begin{array}{c}
{\cal E}\\
{\cal H}
\end{array}\right)=k\left(\begin{array}{cc}
\mathbf{n}\times & 0\\
0 & \mathbf{n}\times
\end{array}\right)\left(\begin{array}{c}
{\cal E}\\
{\cal H}
\end{array}\right),\label{eq:GenEigProblem}
\end{equation}
where $\mathbf{n}$ is the direction of the wave vector and $\mathbf{k}=k\mathbf{n}$.

It is known that the direct eigenproblem corresponds to the Schr\"odinger equation \cite{g101}. 
For the loss-free photonic structure the operator is Hermitian and 
all eigenfrequencies are real. However, in the past several years non-Hermitian 
$\cal{PT}$-symmetric Hamiltonians have 
caught a lot of attention \cite{bender2007making}. An interesting property of 
$\cal{PT}$-symmetric operators is the transition between the completely real 
spectrum and the complex spectrum, which are called exact/spontaneously-broken 
$\cal{PT}$-symmetry. The common situation of the transition is the 
coalescence of two eigenvalues at a branch point, which is referred to as
an exceptional point in the operator theory \cite{kato1966perturbation}. If we 
consider a finite dimensional space and a generalized eigenproblem with 
a perturbed operator in the form $(\hat{L}+\omega\hat{V})\psi = k\hat{M}\psi$ the 
eigenvalues satisfy the characteristic equation having the form of an algebraic equation in 
$\omega$ or $k$. Let us return to Eq.~(\ref{eq:GenEigProblem}). For a 
material with a real dielectric constant we suppose the wave vector $k$ to be a given 
real parameter and write the direct problem with real $\omega(k)$. But now we 
consider $\omega$ to be a given parameter. In the periodic structure for which $\omega$ falls within the allowed frequency bands we have a real $k$, whereas for the forbidden bands the wave vector 
becomes complex. Hence we can regard the bandgap boundaries as the branch points 
and there is an analogy between the band theory and the problem of a spontaneously broken 
$\cal{PT}$-symmetry.

\begin{figure}[!t]
\includegraphics{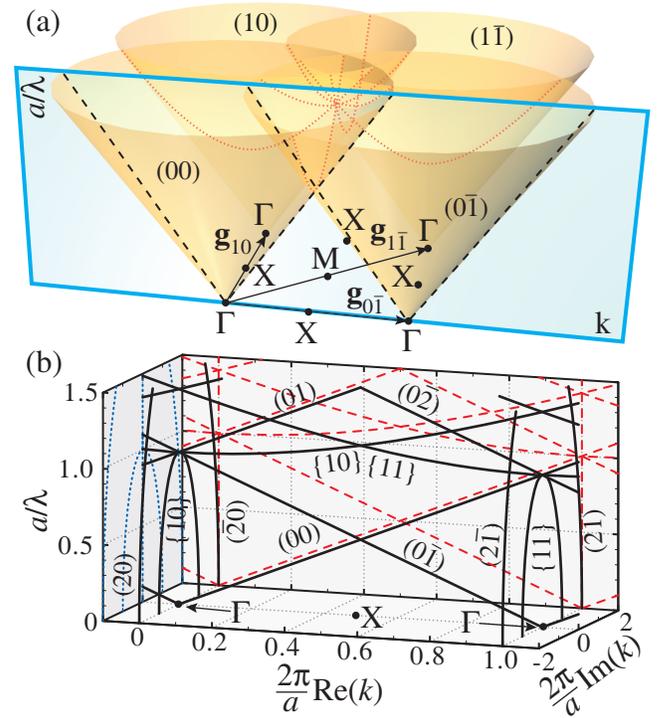}
\caption{
(color online) Complex photonic band diagram for the empty-lattice approximation. (a) Light cone repetitions. The intersections of the cones are marked by the red dotted lines. The cross-section of the cones by the scanning plane $\Gamma\to\mathrm{X}$ are shown by the black dashed lines. (b) Photonic band diagram for complex wave vectors in the $\Gamma\to\mathrm{X}$ direction of the square lattice (black solid lines); its projection on the real (red dashed lines) and imaginary planes (blue dotted lines).
}
\label{fig:EmptyLattice}
\end{figure}

Let us describe a complex band diagram that accounts for both the propagating and evanescent modes.
We start from the empty-lattice approximation (i.e., consider homogeneous 
media) to give a branch notation. For simplicity we consider a two dimensional 
reciprocal space.
For the constant $\cal{E}$ and $\cal{H}$ there is a conical
relation $\omega^2=\varepsilon\mu|\mathbf{k}|^2/c^2$. However the 
periodicity results in that this light cone has many repetitions with the origin at any 
reciprocal lattice vector $\mathbf{g}_{hl}$ [see. Fig.~\ref
{fig:EmptyLattice}(a)].
The Bragg resonances that lead to the scattering from one cone to another 
arise in vicinity of the intersection of the cones, whereas outside the intersection we 
can identify the cones by $(hl)$ indices. As an example we consider a scanning 
plane
in the $\Gamma\to\mathrm{X}$ direction of the Brillouin zone of the square 
lattice. Only the cones $(0l)$ with the origin lying along this direction 
intersect the scanning plane at the low frequency range [Fig.~\ref
{fig:EmptyLattice}(b)]. This is a common situation for the direct $\omega(k)$ method.
If, however, we use the inverse $k(\omega)$ method we obtain the evanescent modes $(hl)$ 
because the cone apex is out of the scanning plane and the conical relation 
gives complex solutions. However in the empty-lattice approximation all the modes 
are orthogonal and there is no coupling
between any modes, and hence the evanescent modes can be neglected, but for the 
case of a nonzero dielectric contrast we cannot do this anymore. Note that in this paper we plot band diagrams in the scale of the dimensionless frequency $a/\lambda$, where $\lambda$ is the wavelength in a vacuum and $a$ is the lattice constant. 
Below we consider the case of the two-dimensional structure however the general case is quite similar. 

\subsection{TE-polarization}

In the TE-polarization we deal with the $z$-component of
the magnetic Bloch's wave $h=\sqrt{\mu_{0}}{\cal H}_{z}$, as well as
with the tangential $e=\sqrt{\varepsilon_{0}}{\cal E}_{\tau}$ and the normal $f=\sqrt{\varepsilon_{0}}{\cal E}_{n}$
components of the electric Bloch's wave. Equation~(\ref{eq:GenEigProblem})
can be rewritten as 
\begin{subequations}
\label{eq:Eq_hef}
\begin{eqnarray}
-i\left(n_{x}\frac{\partial}{\partial x}+n_{y}\frac{\partial}{\partial y}\right) & e(\mathbf{r}) & -i\left(n_{x}\frac{\partial}{\partial y}-n_{y}\frac{\partial}{\partial x}\right)f(\mathbf{r})+\nonumber\\
&+&\frac{\omega}{c}\mu(\mathbf{r})h(\mathbf{r})=-ke(\mathbf{r}),\label{eq:Eq_hefa}
\end{eqnarray}
\begin{equation}
i\left(n_{x}\frac{\partial}{\partial x}+n_{y}\frac{\partial}{\partial y}\right)h(\mathbf{r})-\frac{\omega}{c}\varepsilon(\mathbf{r})e(\mathbf{r})=kh(\mathbf{r}),\label{eq:Eq_hefb}
\end{equation}
\begin{equation}
i\left(n_{x}\frac{\partial}{\partial y}-n_{y}\frac{\partial}{\partial x}\right)h(\mathbf{r})-\frac{\omega}{c}\varepsilon(\mathbf{r})f(\mathbf{r})=0.\label{eq:Eq_hefc}
\end{equation}
\end{subequations}
Note that the functions $h$, $e$ and $f$ are periodic. To proceed further we have 
to reduce these equations to those of the algebraic type. There are two relatively simple  abilities. The first one is to represent operators in plane wave basis, and the second is to use a finite difference scheme in real space. The latter ability is simpler in realization, however it is not work perfectly if the calculating grid is too rare being pure numerical technique
In contrast plane wave basis
makes it possible to evaluate mode properties within two-wave approximations if the modulation of dielectric function is moderate. On the other hand if we consider periodic structure with metal components, plane wave basis is not an optimal choose \cite{li1993convergence,lalanne1996highly}, since zero-order approximation (approximation of free-photons) cannot describe the system correctly, because of free-photon averages dielectric function: positive air with negative dielectric index of metallic inclusions. As a result averaged dielectric index is negative and propagation of light is prohibited that is completely disagree with the correct answer. Hence, to calculate structures with metallic inclusions we should take into account a very large number of plane waves and this technique is worse than finite difference scheme. However, even if we deal with a pure dielectric structure that possesses strong dielectric contrast we have to consider convergence problems as well \cite{sozuer1992convergence}, and the usage of only few number of plane waves is inappropriate. To be sure in correctnesses of our study we examine convergence of traditional $\omega(k)$ plane wave method and find that 625 (25 by 25) plane waves are sufficient to obtain reliable results.

Because the functions $h$, $e$ and $f$ are proportional to the Bloch's amplitudes
they can be represented as Fourier series. We expand $h$, $e$
and $f$ in the r.h.s., find Fourier amplitudes of both sides
of Eqs. (\ref{eq:Eq_hef}), and finally write these equations in the matrix form
\begin{equation}
\mathbf{D}\mathbf{e}+\mathbf{L}\mathbf{f}+\frac{\omega}{c}\mathbf{M}\mathbf{h}=-k\mathbf{e},
\end{equation}
\begin{equation}
-\mathbf{D}\mathbf{h}-\frac{\omega}{c}\mathbf{E}\mathbf{e}=k\mathbf{h},
\end{equation}
\begin{equation}
-\mathbf{L}\mathbf{h}-\frac{\omega}{c}\mathbf{E}\mathbf{f}=0.
\end{equation}
Here $\mathbf{D}$ and $\mathbf{L}$ are diagonal matrices with elements $D_{\mathbf{g},\mathbf{g}}=n_{x}g{}_{x}+n_{y}g{}_{y},$ and
$L_{\mathbf{g},\mathbf{g}}=n_{x}g{}_{y}-n_{y}g{}_{x}$, $\mathbf{E}$
and $\mathbf{M}$ are Toeplitz matrices with elements $E_{\mathbf{g},\mathbf{g}'}=\varepsilon_{\mathbf{g}-\mathbf{g}'}$, and
$M_{\mathbf{g},\mathbf{g}'}=\mu_{\mathbf{g}-\mathbf{g}'}$, which are
the Fourier amplitudes of $\varepsilon(\mathbf{r})$ and $\mu(\mathbf{r})$
respectively. On eliminating the component $\mathbf{f}$, we obtain the conventional
eigenproblem
\begin{equation}
-\left(\begin{array}{cc}
\mathbf{D} & \left(\frac{\omega}{c}\mathbf{M}-\left(\frac{\omega}{c}\right)^{-1}\mathbf{L}\mathbf{E}^{-1}\mathbf{L}\right)\\
\frac{\omega}{c}\mathbf{E} & \mathbf{D}
\end{array}\right)\left(\begin{array}{c}
\mathbf{e}\\
\mathbf{h}
\end{array}\right)=k\left(\begin{array}{c}
\mathbf{e}\\
\mathbf{h}
\end{array}\right).\label{eq:InvDispTE}
\end{equation}

\subsection{TM-polarization}

Following similar procedure we write linear eigenproblem for the electromagnetic field in TM-polarization
\begin{equation}
-\left(\begin{array}{cc}
\mathbf{D} & \left(-\frac{\omega}{c}\mathbf{E}+\left(\frac{\omega}{c}\right)^{-1}\mathbf{L}\mathbf{M}^{-1}\mathbf{L}\right)\\
-\frac{\omega}{c}\mathbf{M} & \mathbf{D}
\end{array}\right)\left(\begin{array}{c}
\mathbf{h}\\
\mathbf{e}
\end{array}\right)=k\left(\begin{array}{c}
\mathbf{h}\\
\mathbf{e}
\end{array}\right).
\label{eq:InvDispTM}
\end{equation} 
Here $\mathbf{h}$ and $\mathbf{e}$ define tangential component of magnetic field and $z$-component of electric field. Matrices $\mathbf{E}$,  $\mathbf{M}$, $\mathbf{D}$, and $\mathbf{L}$ was defined above in TE-polarization subsection. In most cases material does no posses magnetic properties and $\mathbf{M}$ is the identity matrix. Hence Eq.~(\ref{eq:InvDispTM}) is somewhat simpler than Eq.~(\ref{eq:InvDispTE}) since in case of TM polarization the matrix does not contain the inverse of $\mathbf{E}$.


\section{Band diagram for photonic crystals and dielectric metamaterials}

\begin{figure*}[!t]
\includegraphics{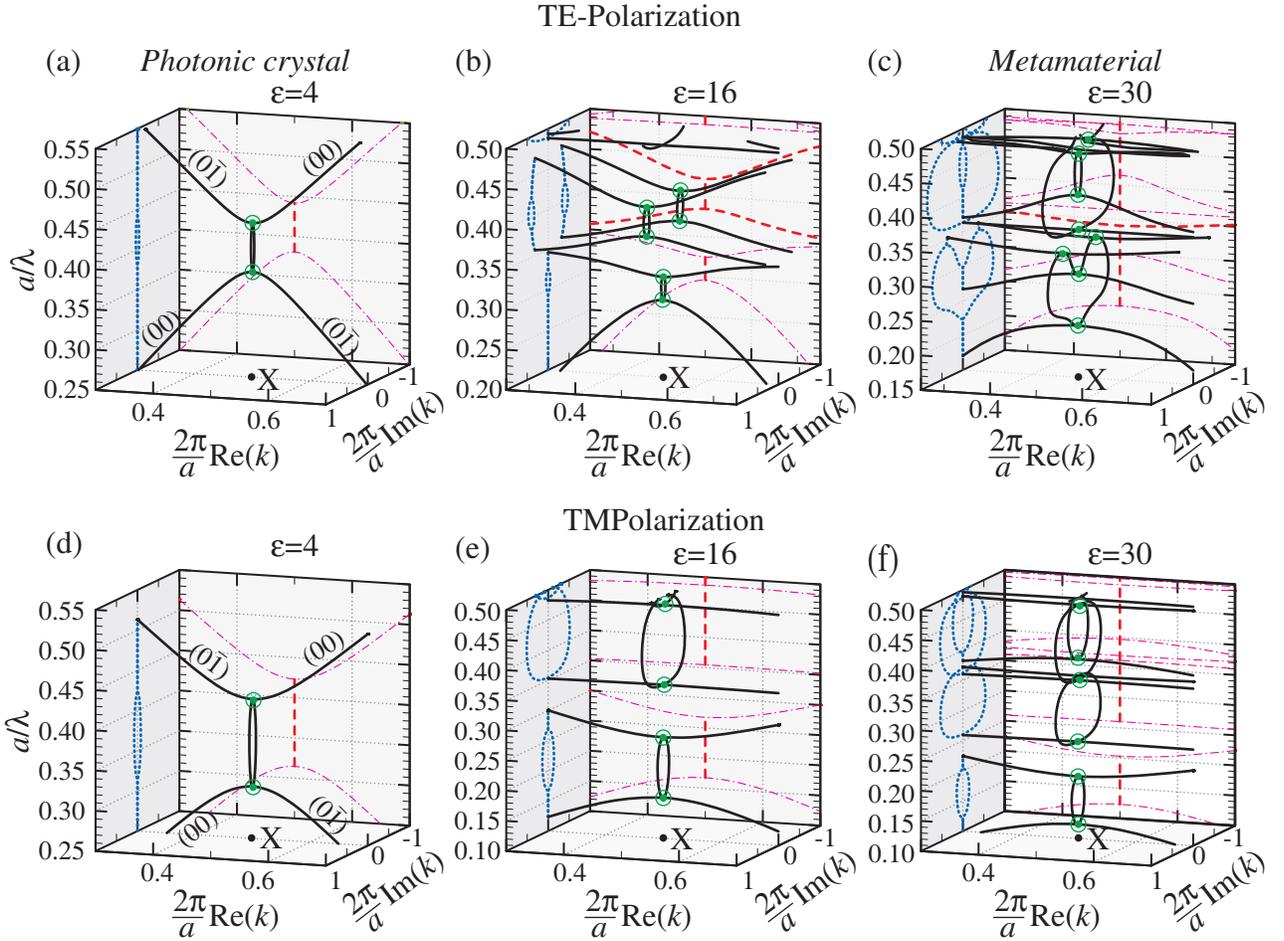}
\caption{
(color online) Complex photonic band diagram for materials with real dielectric constant (a,d) $\varepsilon=4$, (b,e) $\varepsilon=16$, (c,f) $\varepsilon=30$ for (a-c) TE- and (d-f) TM-polarization. The complex wave vector is shown by the black solid curves, and the real part of the wave vector by the dashed curves (real wave vector, thin magenta dash-and-dot curves; complex wave vector, thick red dashed curves). The imaginary part of the wave vector is represented by the blue dotted curves. The branch points are marked by the green circles.
}
\label{fig:RealEps}
\end{figure*}

To illustrate the inverse dispersion method we consider here three types of structures with the same geometry.
Periodic all-dielectric systems are known to be in two scattering regime either as a photonic crystal with the Bragg fundamental gap or as a metamaterial with the Mie fundamental gap \cite{rybin2015phase}. By the fundamental gap we mean the gap between the first and second photonic bands \cite{g101}. Here we consider 2D square lattices of parallel dielectric circular rods infinitely long in the $z$ direction in air with lattice constant $a$ and radius of $r=0.3a$. We study three different dielectric functions $\varepsilon(\omega)$ for the rods corresponding to a-C:H, Si, and Ge$_2$Sb$_2$Te$_5$.

\subsection{Constant dielectric function}

As a reference point we calculated the photonic band diagram by the traditional $\omega(k)$ method in the implementation of Ref. \onlinecite{moroz2011multiple} by using constant values for the dielectric function: $\varepsilon^0_\mathrm{a-C:H}=4$, $\varepsilon^0_\mathrm{Si}=16$, and $\varepsilon^0_\mathrm{GeSbTe}=30$ (the thin magenta  dash-and-dot curves in Fig.~\ref{fig:RealEps}). 
There are bandgaps within the frequency spectra and we have no information about the evanescent modes. However these modes define the scattering, transmission and other light transport properties.
Next, we calculate the photonic band diagram by the inverse dispersion method which yields both propagating and evanescent modes (Fig.~\ref{fig:RealEps}). The bands for the propagating modes (with real wave vector) obtained by using $\omega(k)$ and $k(\omega)$ demonstrate a complete agreement, but now we obtain the dispersion for the evanescent modes.
First we consider TE-polarization.

\subsubsection{TE-polarization}

The spatial modulation of the dielectric function $\varepsilon(r)$ perturbs eigenmodes calculated in the empty-lattice approximation. If the perturbation is small enough (photonic crystal regime) we can identify the branches as $(00)$ and $(0\bar{1})$. Their interaction produces a pair of branch points at the band gap edges marked by the green circles in Fig.~\ref{fig:RealEps}(a,d). 
We describe the frequency dependence of the branches $(00)$ and $(0\bar{1})$. In the low frequency range the wave vector is real. In the interval $0.38 < a/\lambda <0.44$ defined by a pair of the branch points the real part of the wave vector is constant while the imaginary part varies in amplitude, with $\Delta\mathrm{Im}(2\pi k/a)$ of about 0.1. As a result the branches form an narrow elliptical feature (NEF) that is prolate along the frequency axis. The feature corresponds to the bandgap.
With increasing dielectric contrast, the higher frequency modes cannot be adequately identified with the $(hl)$ modes due to the strong interband interaction. In the intermediate case of $\varepsilon=16$ [Fig.~\ref{fig:RealEps}(b)] we can emphasize three NEF: one centered at [$a/\lambda\approx 0.3$, $\mathrm{Im}(2\pi k/a)=0$] and a pair of similar NEF at [$a/\lambda\approx 0.4$, $\mathrm{Im}(2\pi k/a)=\pm 0.5$], which do not intersect the real plane $\mathrm{Im}(2\pi k/a)= 0$.
In the metamaterial regime\cite{rybin2015phase} [Fig.~\ref{fig:RealEps}(c)] we recognize the NEF at [$a/\lambda\approx 0.45$, $\mathrm{Im}(2\pi k/a)=0$]. In addition, we observe two wide features  of complicated shape (WCF). The first WCF is bounded by $0.22<a/\lambda <0.34$ with a width $\Delta\mathrm{Im}(2\pi k/a)\approx 1.6$, and the second lies at $0.37<a/\lambda <0.49$ with a width $\Delta\mathrm{Im}(2\pi k/a)\approx 1.8$.
We notice that all the NEF and WCF in Fig.~\ref{fig:RealEps} originate at the surface of the Brillouin zone (X point). Analysis of the band diagrams reveals that the wave vector of the branch may vary in its real or imaginary part only and these two types of behavior are switched by the branch point.
In addition, the NEF keeps modes connected while the WCF splits the modes to be disconnected at the X point (i.e, we cannot join points at the lower and upper bands).

\subsubsection{TM-polarization}

Now we describe band diagrams for TM-polarization.
When the dielectric contrast is small [Fig.~\ref{fig:RealEps}(d)] the band diagram is similar to the case of TE-polarization [Fig.~\ref{fig:RealEps}(a)]. We identify NEF with parameters $0.31 < a/\lambda <0.42$ and $\Delta\mathrm{Im}(2\pi k/a)\approx 0.15$ being wider than in Fig.~\ref{fig:RealEps}(a)]. Now let us consider band diagram for $\varepsilon=16$, which differs from TE-polarization. Figure~\ref{fig:RealEps}(e) demonstrates just one NEF shifts down $0.17 < a/\lambda <0.28$ and increases its width $\Delta\mathrm{Im}(2\pi k/a)\approx 0.14$ and a WCF ($0.35 < a/\lambda <0.5$ and $\Delta\mathrm{Im}(2\pi k/a)\approx 1.0$). With increasing of dielectric index $\varepsilon$ up to 30 the NEF moves down to the interval $0.12 < a/\lambda <0.20$ ($\Delta\mathrm{Im}(2\pi k/a)\approx 0.3$) and does not disappear (as in TE-polarization) remaining fundamental gap. Also we identify another NEF ($0.40 < a/\lambda <0.49$ and $\Delta\mathrm{Im}(2\pi k/a)\approx 0.5$) and a pair of slightly overlapped WCF ($0.26 < a/\lambda <0.37$, $\Delta\mathrm{Im}(2\pi k/a)\approx 1.07$) and ($0.36 < a/\lambda <0.53$, $\Delta\mathrm{Im}(2\pi k/a)\approx 1.08$). Being similar to the case of TE-polarization the NEF keeps modes connected while the WCF splits the modes to be disconnected at the X point (i.e, we cannot join points at the lower and upper bands).

\begin{figure}[!t]
\includegraphics{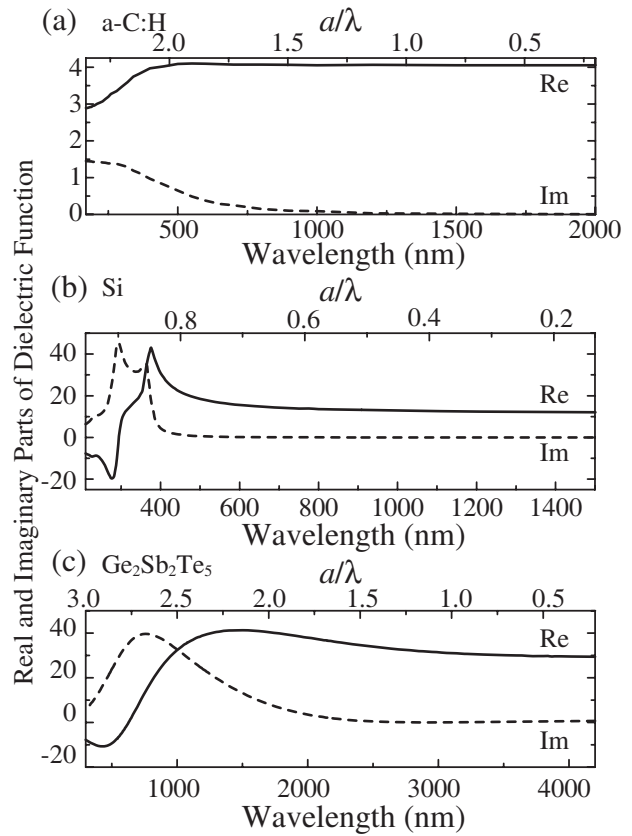}
\caption{
Dielectric functions of (a) a-C:H\cite{palik1998handbook}, (b) Si \cite{aspnes1983dielectric,li1980refractive}, and (c) Ge$_2$Sb$_2$Te$_5$ \cite{nvemec2009ge} in two scales: wavelength $\lambda$ and $a/\lambda$ for lattice constants (a) $a=400$ nm, (b) $a=200$ nm, and (c) $a=900$ nm. The real part is shown by the solid curve, and the imaginary part by the dashed curves.  
}
\label{fig:DielFunct}
\end{figure}

\begin{figure*}[!t]
\includegraphics{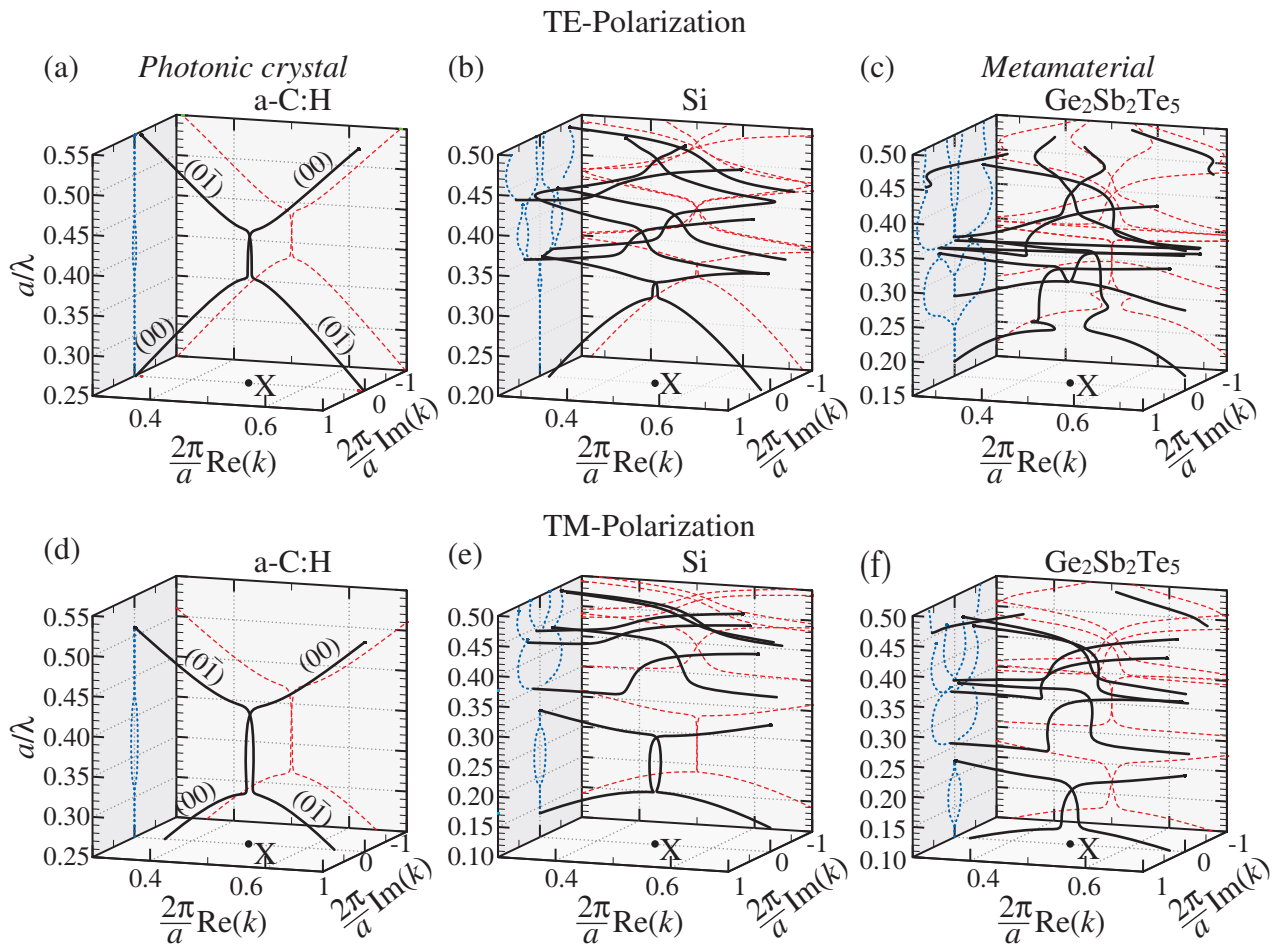}
\caption{
(color online) Complex photonic band diagram for the materials with the experimentally measured dielectric functions: (a,d) a-C:H \cite{palik1998handbook}, (b,e) Si \cite{aspnes1983dielectric,li1980refractive}, (c,f) Ge$_2$Sb$_2$Te$_5$ \cite{nvemec2009ge} for (a-c) TE- and (d-f) TM-polarization. The complex wave vector is shown by the black solid curves; the real part of the wave vector by the red dashed curves; the imaginary part of the wave vector by the blue dotted curves.
}
\label{fig:MaterialEps}
\end{figure*}

\subsection{Experimental dielectric function}

We can calculate photonic band diagrams for the three systems under consideration by using the experimentally measured dielectric response of a material \cite{li1980refractive,aspnes1983dielectric,palik1998handbook,nvemec2009ge} plotted in Fig.~\ref{fig:DielFunct}, since the inverse dispersion method is applicable for calculation of an arbitrary dielectric function. 
As far as Maxwell's equations are not frequency-scalable for non-constant dielectric functions we have to specify  both the material and the geometrical sizes of the 2D structures: (i) for a-C:H the lattice constant is set to $a=400$ nm and the dielectric function is taken from Ref. \onlinecite{palik1998handbook}; (ii) for Si the lattice constant is set to $a=200$ nm, the dielectric function is taken from Refs.~\onlinecite{li1980refractive,aspnes1983dielectric}; and (iii) $a=900$ nm for a metamaterial fabricated form Ge$_2$Sb$_2$Te$_5$ (fcc phase of a GeSbTe chalcogenide glass)\cite{nvemec2009ge}. Calculated photonic band diagrams are presented in Fig.~\ref{fig:MaterialEps} and comparison with Fig.~\ref{fig:RealEps} reveals a number of essential differences in the mode behavior. The imaginary part of the dielectric function lifts the degeneracy at the branch points marked by the green circles in Fig.~\ref{fig:RealEps}. It is convenient to introduce the density of states per interval of a real part of wave vector $\rho(k')$ for the branches. In the case of real dielectric function both types of features NEF and WCF lie at the surface of the Brillouin zone [$\mathrm{Re}(\mathbf{k})=\mathrm{X}$] with the density of states $\rho(k')$ becoming a delta-function at the X point. With  increasing imaginary part of the dielectric function, the density of states broadens away from the surface of the Brillouin zone.

\begin{figure}[!t]
\includegraphics{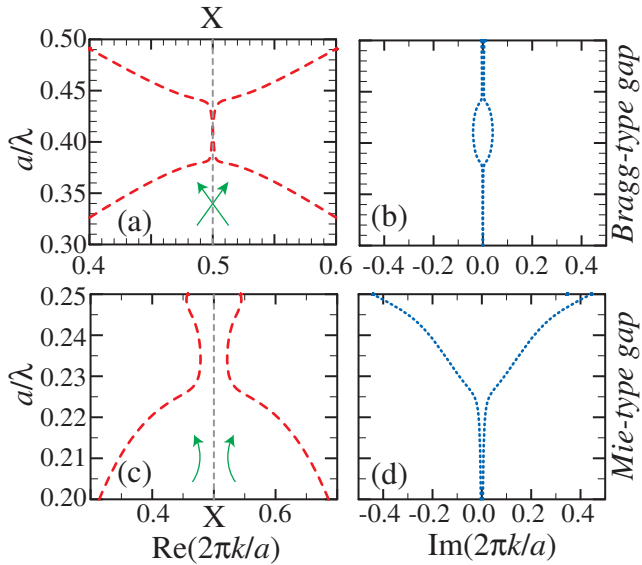}
\caption{
(color online) Comparison of the Bragg-type gap (structure composed of a-C:H) and the Mie-type gap (structure composed of Ge$_2$Sb$_2$Te$_5$) in TE-polarization. The band diagram is for the real (a) and imaginary (b) parts of the wave vector in the photonic crystal scattering regime and for the real (c) and imaginary (d) parts of the wave vector in the Mie scattering regime. The green arrows demonstrate the behavior of the branches at the boundary of the Brillouin zone marked by the gray dashed line.
}
\label{fig:BraggMie}
\end{figure}

Also, there is a difference in the mode behavior at NEF and WCF when it is fundamental gap. We notice that WCF becomes fundamental gap only in TE-polarization. Figure~\ref{fig:BraggMie} shows the band gaps related to different features NEF and WCF on a detailed scale. The group velocity $V=(d\mathrm{Re}(k)/d\omega)^{-1}$ does not change its direction at NEF, while at WCF the group velocity $V_\mathrm{WCF}$ becomes zero and changes its direction to the opposite. By contrast, in the case of a real dielectric function there is an interval of a vanished group velocity hence the dielectric function needs to possess at least a small amount of an imaginary part to distinguish the type of the gap.

\section{Discussion}

The inverse dispersion method is applicable for accurate calculating of the photonic band diagram for periodic structures composed of materials with an arbitrary dielectric response whereas the conventional techniques cannot describe the photonic properties correctly. Also the method provides a deep insight 
into the physics of propagation of traveling of electromagnetic waves in periodic media including photonic crystals and metamaterials. 
The photonic band diagram calculated by the direct $\omega(k)$ method hides the branch points because this method is inapplicable to finding of the evanescent modes. At the same time the linear operator theory describes singular properties of eigenmodes at the branch point \cite{kato1966perturbation}.

In the imaginary plane of the band diagram, the NEF is unambiguously associated with the Bragg resonance, because it increases at very small dielectric contrast and involves a pair of modes [Fig.~\ref{fig:BraggMie}(b)]. The wide complicated feature is associated with the Mie resonance because of being related to the Mie gap in all-dielectric structures as shown in Ref. \onlinecite{o2002photonic}. The complex photonic band diagram in Fig.~\ref{fig:RealEps}(c) demonstrates that the Mie resonance feature extends from $a/\lambda=0.22$ to $0.35$ (see in the imaginary plane) being much wider than the Mie-band gap $0.22 \leqslant a/\lambda \leqslant 0.26$ (thin dash-and-dot curves in the real plane). Analysis of the frequency ranges of the resonant Mie modes \cite{rybin2013mieOE,rybin2015switching} shows that the lower Mie feature in Fig.~\ref{fig:RealEps}(c) is associated with the TE$_{01}$ Mie mode while the higher feature is relater to the TE$_{11}$ mode. In a similar way WCF in Fig.~\ref{fig:RealEps}(c) are originated from the TM$_{01}$ and TM$_{11}$ Mie modes.  
We also demonstrate that the Mie-related WCF make the band diagram disconnected at the X point [Fig.~\ref{fig:RealEps}(c)].
 
Turning to the Eqs.~(\ref{eq:InvDispTE})~and~(\ref{eq:InvDispTM}), we notice that they are non-Hermitian operators with the wave vector length as an eigenvalue. Therefore we can consider these operators to be a wave vector operator for the periodic media. Inspired by the results obtained from non-Hermitian $\cal{PT}$-symmetric Hamiltonians \cite{bender2007making}
 we assume that the wave vector operator possesses a certain symmetry. The symmetry 
$\cal{T}$ is related to the time-reversal symmetry, while the parity 
$\cal{P}$ should be applied to the fields represented in the $k$-space (we notice that the real space representation is no better than the $k$-space representation). The important consequence of the description of operator (\ref{eq:InvDispTE})~and~(\ref{eq:InvDispTM}) as operators with $\cal{PT}$-symmetry is that we should not consider these operators as a mathematical trick only and reject they as being unphysical because they violate the axiom of Hermiticity.
Conversely we do consider these operators as physical quantities that are related to momentum of excitation in periodical media. At the branch point the symmetry changes from exact to spontaneously-broken that makes this excitation prohibited.

In addition, using the results of the inverse dispersion method we can separate the Bragg and Mie gaps by introducing a small amount of loss. The modes forming Bragg gaps do not change their direction, while modes forming the Mie gaps show a repulsive behavior. Using this criteria we  explicitly identify the lower bandgap as a Bragg gap (Fig.~\ref{fig:BraggMie}).

\section{Conclusion}

We have suggested the inverse dispersion method for calculating the complex photonic band diagram for periodic media with an arbitrary dielectric function. The method makes it possible to calculate both the propagating and evanescent modes. We have calculated the band diagram of periodic structures composed of the following materials: low-refractive index amorphous hydrogenated carbon (constituting a photonic crystal), high-refractive index silicon, and chalcogenide glass (constituting an all-dielectric metamaterial) by using the experimentally measured properties of the material. 
Our analysis demonstrated the essential difference between the fundamental gaps that originates from Bragg or Mie evanescent modes. This result is of fundamental importance for studies of the light scattering regimes of photonic crystals and metamaterials\cite{rybin2015phase}. Moreover we note that the inverse dispersion method is not restricted to only photonic diagrams, being rather applicable to any waves in periodic media: electrons, phonons, magnons etc.

\begin{acknowledgments}
{ We acknowledge fruitful discussions with A.A.~Kaplyanskii, Yu.S. Kivshar, S.F. Mingaleev, and K.B.~Samusev. This work was supported by the Russian Foundation for Basic Research (15-02-07529).
}
\end{acknowledgments}



\end{document}